\documentclass[sigconf,nonacm]{acmart}

\usepackage{amsmath}
\usepackage{algorithm}
\usepackage{algpseudocode}
\usepackage{listings}
\usepackage{xcolor}
\usepackage{graphicx}
\usepackage{booktabs}
\usepackage{multirow}
\usepackage{subcaption}
\usepackage{tikz}
\usepackage{float}
\usepackage{pgfplots}
\usepackage{placeins}
\usepackage{makecell}
\usepackage{nicefrac}
\usepackage{textcomp}
\usepackage{siunitx}
\usepgfplotslibrary{groupplots}
\usetikzlibrary{arrows.meta, positioning, calc, decorations.pathreplacing}

\lstdefinelanguage{MLIR}{
  keywords={func, air, launch, segment, herd, channel, put, get, token, memref, affine, for, async, dma},
  keywordstyle=\color{blue}\bfseries,
  comment=[l]{//},
  commentstyle=\color{gray}\itshape,
  stringstyle=\color{red},
  basicstyle=\ttfamily\small,
  breaklines=true,
  frame=single,
  showstringspaces=false
}

\algnewcommand{\algorithmicparallel}{\textbf{parallel for}}
\algdef{S}[PARFOR]{ParFor}[1]{\algorithmicparallel\ #1\ \textbf{do}}
\algdef{E}[PARFOR]{EndParFor}{\textbf{end parallel for}}

\begin{document}

\title{Programming AMD XDNA\texttrademark\ NPUs with Open-source\\Compiler Tools: A FlashAttention Case Study}

\author{%
  Erwei~Wang\textsuperscript{*},
  Ephrem~Wu\textsuperscript{*},
  Victor~J.~B.~Jung\textsuperscript{\dag},
  Jiajie~Li\textsuperscript{\ddag},
  Andr\'{e}~R\"{o}sti\textsuperscript{*},
  Joseph~Melber\textsuperscript{*},
  Samuel~Bayliss\textsuperscript{*}%
}
\affiliation{%
  \institution{%
    \textsuperscript{*}Advanced Micro Devices, Inc., USA.\\
    \textsuperscript{\dag}Integrated Systems Laboratory (IIS), ETH Z\"urich, Switzerland.\\
    \textsuperscript{\ddag}Cornell University, Ithaca, NY, USA.%
  }
  \country{}%
}
\renewcommand{\shortauthors}{Wang et al.}
\makeatletter
\def\@authorfont{\Large}
\makeatother

\begin{abstract}
Spatial NPUs such as AMD XDNA\texttrademark\ place compute tiles beside small
local memories and leave data movement between them to software. Mapping a
multi-stage workload onto such a device is largely a question of where the
intermediate tensors live. We report what we learned making those choices for
FlashAttention with the open-source IRON and MLIR-AIR flows.

We compare four reference designs on XDNA~1 and XDNA~2: one runs each operator
separately, two stream between operators on chip, and one fuses all three
attention stages into a single kernel. The fused kernel holds the
$\boldsymbol{QK}^{\mathsf T}$ scores in compute-tile local memory and reduces
partial results over the cascade interconnect, so the scores never return to
shared MemTile memory. On XDNA~2, it reaches 3.62~TFLOP\textsubscript{GEMM}/s over
complete end-to-end execution, twice the IRON design, with 5.3 to 7.2 times the
energy efficiency of the integrated GPU on the same chip at 2K tokens and above.
It covers twelve LLM configurations, from BERT to DeepSeek, up to 128K tokens.

Roofline analysis at each memory level explains this result and shows when to
stop. XDNA~1 has lower ridge points, so streaming on chip already reaches the
compute-bound regime: the same fusion that doubles throughput on XDNA~2 is nearly
wasted on XDNA~1. Comparing a mapping's operational intensity against each level's
ridge point predicts which case applies before writing any code. Fuse until the
mapping clears that ridge point, then stop. We release the reference designs as
maintained open source.
\end{abstract}



\maketitle

\section{Introduction}
\label{sec:intro}

Spatial neural processing units (NPUs) are increasingly important for edge AI inference because they offer much higher energy efficiency than general-purpose processors under tight power and thermal budgets. Their appeal is particularly strong for transformer inference at the edge, where latency, privacy, and dependence on cloud connectivity all favor local execution.
At the same time, programming NPUs efficiently is challenging. Unlike CPUs and GPUs, spatial NPUs typically expose distributed memories, explicit data movement, and architecture-specific communication fabrics, so performance depends heavily on how computation is partitioned and mapped onto the array.

The AMD XDNA\texttrademark\ architecture\cite{ricoXDNAnpu}, implemented in the NPU of AMD Ryzen\texttrademark\ AI processors, is an example of a spatial architecture. It comprises compute tiles, memory tiles (MemTiles), and shim tiles. Each compute tile contains local scratchpad memory together with scalar and vector engines, while each MemTile contains bigger scratchpad memory acting as shared memory, but without compute engines. All tile types provide hardware DMA engines that connect tile-local memories to programmable streaming interconnects. Additional neighbor-to-neighbor cascade connections along rows or columns provide extra bandwidth for tightly coupled data movement and reductions. Realizing peak throughput for complex, memory-intensive operators therefore requires careful co-design of algorithm, data movement, and hardware mapping.

Scaled dot-product attention~\cite{vaswani2017attention}, the computational core
of transformer models, is hard to map onto a spatial NPU for two reasons. The
score intermediate $\mathbf{Q}\mathbf{K}^{\mathsf T}$ is large, growing with the
square of the sequence length, so where it lives dominates the memory traffic.
Separately, $\mathbf{Q}$, $\mathbf{K}$, and $\mathbf{V}$ each call for a different
reuse and broadcast pattern, yet a compute tile on XDNA has only two inbound DMA
channels to deliver them. Those channels must be time-multiplexed, and a single
mapping has to satisfy both constraints at once.

FlashAttention~\cite{dao2022flashattention} answers the first constraint on GPUs
by tiling the computation so the scores stay in shared memory. That answer does
not carry over unchanged. A GPU thread block pulls its data through a cache
hierarchy, whereas on a spatial NPU the program places the data itself, in
distributed local memories reached over DMA engines and programmable streaming
interconnects.

A key design question is \emph{where} the bandwidth demand associated with the $\mathbf{Q}\mathbf{K}^{\mathsf T}$ intermediate is absorbed within the memory hierarchy. Mapping strategies differ primarily in whether the $\mathbf{Q}\mathbf{K}^{\mathsf T}$ scores are materialized in DDR, spilled through shared MemTile memory, or kept entirely within compute tile-local memory.
This choice is a dominant factor in determining the resulting
performance.

Several NPU mapping strategies have been proposed---from simple layer-by-layer execution to pipelined on-chip dataflow to fully fused kernel execution.
Prior DATO~\cite{fang2025dato} and IRON~\cite{hunhoff2025iron} reference designs provide two concrete on-chip attention mappings with different stage organizations and MemTile communication patterns. Together with the layer-by-layer control and fused MLIR-AIR~\cite{wang2025mlir-air} design, they illustrate a practical progression toward more integrated spatial execution.
However, no systematic study has compared these strategies on the same NPU hardware and analyzed \emph{why} each successive step in this design space improves performance. Without such analysis, practitioners lack actionable guidance for mapping similar workloads.

This paper fills that gap. We present a four-point ablation study on AMD XDNA\texttrademark ~1 and XDNA\texttrademark ~2 NPUs using FlashAttention as a case study, progressing from naive to fully optimized:

\begin{enumerate}
\item \textbf{Layer-by-layer}: Each operation (QK matmul, causal masking, softmax, GV matmul) runs as a separate kernel. Data returns to DDR between stages, and the device is fully reconfigured between them.
\item \textbf{Streamed dataflow (DATO)}: A pipelined on-chip dataflow implemented using the DATO programming abstraction that connects stages via streams, eliminating DDR round-trips but retaining intermediate data spilling at MemTile memory. No device reconfiguration is required.
\item \textbf{Streamed dataflow (IRON)}: A different pipelined on-chip dataflow with reduced MemTile memory spilling, implemented using the IRON programming abstraction.
\item \textbf{Fused kernel (MLIR-AIR)}: All three stages execute on every tile. Intermediates stay in compute-tile memory, and cross-tile reduction uses the hardware cascade. Implemented using the MLIR-AIR programming abstraction.
\end{enumerate}

We analyze each transition through roofline modeling at three memory hierarchy levels, revealing how the performance bottleneck shifts across the hierarchy as optimization progresses: from DDR bandwidth (layer-by-layer), to on-chip bandwidth (pipelined dataflow), to compute-bound execution (fused kernel). We demonstrate generality and quality of results across LLM architectures and validate functional correctness through numerical stability analysis and Llama end-to-end integration.

The main contributions of this work are as follows.

\begin{itemize}

    \item We show how open-source compiler tools encode hardware-software co-design knowledge into reusable automation, reducing low-level NPU programming effort while delivering a fused FlashAttention reference design that reaches 3.62 effective-matmul TFLOP/s (TFLOP\textsubscript{GEMM}/s), measured over the complete end-to-end FlashAttention execution, on XDNA~2.

    \item We instantiate and evaluate this methodology through a systematic ablation study on AMD XDNA NPUs, using attention as a case study. The study compares layer-by-layer execution, DATO, IRON, and MLIR-AIR, and relates their performance differences to the changing bottleneck across the memory hierarchy.

    \item The practicality of this methodology is demonstrated by developing a fused attention mapping for AMD NPUs that reaches 3.62~TFLOP/s of GEMM-only throughput and 6.5$\times$ energy efficiency relative to the integrated GPU (iGPU) on the same device on XDNA~2, generalized across representative LLM configurations including BERT, GPT-2/OPT, Llama, Qwen, and DeepSeek.
\end{itemize}

Rather than proposing these techniques individually, this paper is intended to share the practical experience needed to apply them together effectively on XDNA devices.

\section{Background}
\label{sec:background}

In this section, we first review attention and FlashAttention to identify the computation structure, dominant intermediates, and candidate tiling dimensions. We then summarize the AMD XDNA architectural features and programming abstractions that shape how programmers can map these workloads onto the NPUs with improved efficiency.

\subsection{Attention and FlashAttention}
\label{sec:flashattn}

For a single attention head, scaled dot-product attention is
\begin{equation}
\mathrm{Attention}(\boldsymbol{Q}, \boldsymbol{K}, \boldsymbol{V})
=
\mathrm{softmax}\!\left(\frac{\boldsymbol{Q}\boldsymbol{K}^{\mathsf T}}{\sqrt{d_k}}\right)\boldsymbol{V},
\end{equation}
where $\boldsymbol{Q} \in \mathbb{R}^{l_q \times d_k}$, $\boldsymbol{K} \in \mathbb{R}^{l_k \times d_k}$, $\boldsymbol{V} \in \mathbb{R}^{l_k \times d_v}$, and the output in $\mathbb{R}^{l_q \times d_v}$. In multi-head attention, this computation is replicated across $H$ heads.

These dimensions expose several tiling candidates. The head dimension $H$ is embarrassingly parallel. The query dimension $l_q$ and output feature dimension $d_v$ are also naturally parallel, as distinct query rows and output channels can be computed independently once the softmax statistics are known. In contrast, the key/value sequence dimension $l_k$ is a reduction dimension: each query row must aggregate contributions from all keys, and the feature dimension $d_k$ is an inner-product reduction in the score computation $\boldsymbol{Q}\boldsymbol{K}^{\mathsf T}$. Thus, $H$, $l_q$, and $d_v$ primarily expose parallelism, whereas $l_k$ and $d_k$ induce reductions.

\begin{algorithm}[h]
\caption{Per-tile FlashAttention.}
\label{alg:cascade-stage}
\begin{algorithmic}[1]
\Procedure{FLASHATTENTION}{$\boldsymbol{Q}^\prime, \boldsymbol{K}, \boldsymbol{V}, l_k^\prime$}
    \State $\triangleright\ \boldsymbol{Q}^\prime \in \mathbb{R}^{l_q^\prime \times d_k}, \boldsymbol{K} \in \mathbb{R}^{l_k \times d_k}, \boldsymbol{V} \in \mathbb{R}^{l_k \times d_v}, l_k^\prime|l_k$
    \State $\boldsymbol{A} \leftarrow \nicefrac{\boldsymbol{Q}^\prime }{\sqrt{d_k}}$
    \State $\boldsymbol{G}^\prime \leftarrow 0$
    \State $(\boldsymbol{u}', \boldsymbol{s}') \leftarrow (-\infty, 0)$
    \For{$j \in \left[0, \nicefrac{l_k}{l_k^\prime}\right)$}
        \If{causal mode}
            \State $\boldsymbol{G} \leftarrow$ \Call{ApplyCausalMask}{}
        \EndIf
        \State $\boldsymbol{B} \leftarrow \boldsymbol{K}\left[jl_k^\prime : (j+1)l_k^\prime, :\right]$
        \State $\boldsymbol{G} \leftarrow \boldsymbol{A} \cdot \boldsymbol{B}^{\mathsf T} + \boldsymbol{G}$
        \State $\boldsymbol{u} \leftarrow \max_{z}(\boldsymbol{G}:z)$
        \State $\boldsymbol{u} \leftarrow \max_{z}(\boldsymbol{u}^\prime, \boldsymbol{u})$
        \State $\boldsymbol{G} \leftarrow \exp(\boldsymbol{G} - \boldsymbol{u} \cdot \boldsymbol{1}^{\mathsf T})$
        \State $\boldsymbol{G} \leftarrow \mathrm{convert}(\boldsymbol{G},\text{ bfloat16})$
        \State $\boldsymbol{B} \leftarrow \boldsymbol{V}[jl_k^\prime : (j+1)l_k^\prime]$
        \State $\boldsymbol{r} \leftarrow \exp(\boldsymbol{u}' - \boldsymbol{u})$
        \State $\boldsymbol{G}^\prime \leftarrow \boldsymbol{G} \cdot \boldsymbol{B} + \boldsymbol{G}^\prime \odot \boldsymbol{r} \cdot \boldsymbol{1}^{\mathsf T}$
        \State $\boldsymbol{s} \leftarrow \sum_{z} \boldsymbol{G}\left[:,z\right]$
        \State $\boldsymbol{s} \leftarrow \boldsymbol{s} + \boldsymbol{s}^\prime \odot \boldsymbol{r}$
        \State $\left(\boldsymbol{u}^\prime, \boldsymbol{s}^\prime\right) \leftarrow \left(\boldsymbol{u}, \boldsymbol{s}\right)$
    \EndFor
    \State \Return $\left(\boldsymbol{G}^\prime, \boldsymbol{u}, \boldsymbol{s}\right)$
\EndProcedure
\end{algorithmic}
\end{algorithm}

\begin{algorithm}[h]
\caption{Tile-to-tile FlashAttention}
\label{alg:cascade-merge}
\begin{algorithmic}[1]
\Procedure{SPATIAL\_FLASHATTENTION}{$\boldsymbol{Q}, \boldsymbol{K}, \boldsymbol{V}, l_q^\prime, l_k^\prime, C$}
    \State $\triangleright\ \boldsymbol{Q} \in \mathbb{R}^{l_q \times d_k}, \boldsymbol{K} \in \mathbb{R}^{l_k \times d_k}, \boldsymbol{V} \in \mathbb{R}^{l_k \times d_v}, l_q^\prime | l_q, C | l_k, \nicefrac{l_k^\prime | l_k}{C}$.
    \For{$i \in \left[0,\, l_q/l_q^\prime\right)$}
        \State $\boldsymbol{Q}^\prime \leftarrow \boldsymbol{Q}[i l_q^\prime : (i+1) l_q^\prime, :]$
        \ParFor {$c \in [0,\, C)$}
            \State $\mathrm{row\_slice} \equiv \nicefrac{c l_k}{C} : \nicefrac{(c+1)l_k}{C}$
            \State $(\boldsymbol{G}^\prime[c], \boldsymbol{u}^\prime[c], \boldsymbol{s}^\prime[c]) \leftarrow \mathrm{FLASHATTENTION} (\boldsymbol{Q}^\prime,$
            \Statex \hspace{\algorithmicindent} \hspace{\algorithmicindent} \hspace{\algorithmicindent} \hspace{\algorithmicindent} $\boldsymbol{K}[\mathrm{row\_slice},:],\, \boldsymbol{V}[\mathrm{row\_slice},:],\, l_k^\prime)$
        \EndParFor
        \State $(\boldsymbol{G}, \boldsymbol{u}, \boldsymbol{s}) \leftarrow (\boldsymbol{G}^\prime[0], \boldsymbol{u}^\prime[0], \boldsymbol{s}^\prime[0])$
        \For{$c \leftarrow 1,2,\dots,C-1$}
            \State $\boldsymbol{u} \leftarrow \max(\boldsymbol{u}, \boldsymbol{u}^\prime[c])$
            \State $\boldsymbol{r} \leftarrow \exp(\boldsymbol{u}^\prime[c-1] - \boldsymbol{u})$
            \State $\boldsymbol{G} \leftarrow \boldsymbol{G}^\prime[c]  + \boldsymbol{G} \odot \boldsymbol{r}$
            \State $\boldsymbol{s} \leftarrow \boldsymbol{s}^\prime [c] + \boldsymbol{s} \odot \boldsymbol{r}$
        \EndFor
    \EndFor
    \State \Return $\boldsymbol{G} \odot \left(\nicefrac{1}{\left(\boldsymbol{s} \cdot \boldsymbol{1}^{\mathsf T}\right)}\right)$
\EndProcedure
\end{algorithmic}
\end{algorithm}

A naive implementation materializes the full score matrix in $\mathbb{R}^{l_q \times l_k}$, incurring quadratic memory traffic that dominates for long sequences. FlashAttention~\cite{dao2022flashattention} addresses this by tiling the reduction over $l_k$: it streams blocks of $\boldsymbol{K}$ and $\boldsymbol{V}$ through fast memory while maintaining online softmax statistics (row-wise max and sum).~\cite{milakov2018online,rabe2021selfattention} This avoids materializing the full attention matrix and writes only the final output, achieving IO-optimized memory complexity.~\cite{dao2022flashattention}

FlashAttention maps naturally to spatial accelerators via a two-level decomposition: \textit{temporal} tiling over the sequence dimension, and \textit{spatial} tiling across parallel tiles.

Each tile processes a query block $\boldsymbol{Q}' \in \mathbb{R}^{l_q \times d_k}$ against streamed $\boldsymbol{K}/\boldsymbol{V}$ blocks $\boldsymbol{K}_j \in \mathbb{R}^{l_k \times d_k}$ and $\boldsymbol{V}_j \in \mathbb{R}^{l_k \times d_v}$. For each query row, the tile maintains the running max $\boldsymbol{u}$, normalization sum $\boldsymbol{s}$, and unnormalized output accumulator $\boldsymbol{G}'$.
After all $\boldsymbol{K}/\boldsymbol{V}$ blocks are processed, the output block is $\boldsymbol{G}' / \boldsymbol{s}$. This corresponds to Algorithm~\ref{alg:cascade-stage}.

Looking at the tile array, the context is partitioned across $C$ spatial tiles. Each tile computes a partial summary $(\boldsymbol{G}'_c, \boldsymbol{u}_c, \boldsymbol{s}_c)$ over its assigned $\boldsymbol{K}/\boldsymbol{V}$ range.
After reduction across all tiles, the final output is $\boldsymbol{O}' = \boldsymbol{G}'_{\mathrm{agg}} / \boldsymbol{s}_{\mathrm{agg}}$, as shown in Algorithm~\ref{alg:cascade-merge}.

There are multiple ways to map FlashAttention onto a spatial accelerator. One option is a purely temporal mapping, where each tile streams $\boldsymbol{K}/\boldsymbol{V}$ blocks sequentially and performs the full online softmax reduction locally. Another is a hierarchical mapping that exploits both temporal and spatial decomposition: the inner level reduces over temporally streamed $\boldsymbol{K}/\boldsymbol{V}$ tiles, while the outer level merges partial results produced across spatially distributed tiles.

\subsection{AMD XDNA Architecture}
\label{sec:aie2p}

We evaluate on two generations of AMD XDNA NPUs: Ryzen AI 7040/8040 Series processor's NPU, based on the AMD XDNA~1 architecture, and Ryzen AI 300/400 Series processor's NPU, based on the AMD XDNA~2 architecture~\cite{amd-ryzen-ai}. Both organize the NPU as a 2D array of tiles, where each column contains a stack of compute tiles, a MemTile, and a shim tile.
In Figure~\ref{fig:dataflow}, each vertical stack corresponds to one NPU column, comprising four compute tiles, a MemTile, and a shim tile.
XDNA~1 provides a 4$\times$4 compute-tile array, whereas XDNA~2 widens to 4$\times$8. Both architectures provide private scratchpad in each compute tile, shared MemTile memory, and hardware cascade connections between neighboring compute tiles, but differ in array width, clock speed, MAC throughput, and transcendental support.
A detailed summary of their configurations is shown in Table~\ref{tab:hw-comparison}.

Programming XDNA requires a different execution model from conventional load/store architectures. CPU and GPU programs are largely pull-oriented: executing threads issue loads, and the memory hierarchy dynamically supplies the requested data. XDNA instead exposes a \textit{push-oriented} spatial model. The program configures a network of compute tiles, local buffers, DMA engines, and stream routes. The DMA engines then push data into distributed memories, and tile programs consume and produce values according to explicit producer-consumer schedules. The compiler or programmer must therefore determine not only what each processing element computes, but also where and when it executes, how its operands reach it, and how buffering and synchronization coordinate computation with data movement. This explicit organization can avoid the overhead of a general-purpose cache hierarchy, but performance depends on matching compute rates, DMA schedules, routes, and buffer capacities.

\begin{table}[t]
\caption{XDNA~1 and XDNA~2 hardware configurations used in this paper. They reflect only the evaluated experimental settings.}
\label{tab:hw-comparison}
\centering
\small
\begin{tabular}{@{}lll@{}}
\toprule
\textbf{Feature} & \textbf{XDNA~1} & \textbf{XDNA~2} \\
\midrule
Compute tile array & 4$\times$4 (16 tiles) & 4$\times$8 (32 tiles) \\
Clock frequency & 1.0 GHz & 1.3 GHz\footnotemark[1] \\
MAC data type & \texttt{bfloat16} & \texttt{bfp16}\footnotemark[2] \\
MAC per clk. ($m\times k\times n$) & 4$\times$8$\times$4 & 8$\times$8$\times$8 \\
Exp. intrinsic & LUT approximation & Vector \texttt{exp2} \\
Compute tile memory & 64 KB & 64 KB \\
Memtile memory & 512 KB & 512 KB \\
\bottomrule
\end{tabular}
\end{table}

\footnotetext[1]{Variable. Fixed at 1.3 GHz in this paper.}
\footnotetext[2]{\texttt{bfp16} means block floating point. Data I/O uses \texttt{bfloat16}, and values are vector-cast before computation on XDNA~2.}

\subsection{Programming Abstractions for AMD XDNA NPU}
\label{sec:programming-abstractions}

The programming systems used in this study provide different abstractions for expressing the same responsibilities of a spatial NPU mapping: compute placement, data placement, communication, temporal scheduling, synchronization, and interconnect selection. These controls are performance-critical on XDNA because data is explicitly pushed between distributed memories and compute tiles. A high-quality mapping must coordinate compute programs and DMA engines as one spatial dataflow rather than treating data movement as a secondary implementation detail.

IRON~\cite{hunhoff2025iron} provides the closest-to-metal interface among the three. It builds directly on the MLIR-AIE infrastructure and exposes the spatial structure of the XDNA array, including compute tiles, MemTiles, shim tiles, and DMA-managed communication. Programs are written as explicit device designs: compute kernels are assigned to tiles, communication is represented through ObjectFifos and DMA tasks, and host-side runtime sequences configure and launch asynchronous transfers. This makes IRON well suited for hand-optimized FlashAttention implementations where the programmer directly controls tiling, double buffering, data layout, and the placement of pipeline stages across the NPU fabric.

DATO~\cite{fang2025dato} raises the programming model to a task-based dataflow abstraction. Rather than constructing individual tile programs and DMA sequences directly, the programmer expresses an application as a graph of tasks connected by typed streams. DATO makes communication and sharding first-class program constructs through \texttt{Stream} and \texttt{Layout} types, allowing producer--consumer relationships and tensor partitioning to be stated explicitly in the source program. For FlashAttention, this abstraction naturally captures the algorithm as a streaming pipeline of load, matrix multiply, softmax, rescaling, and output stages, while delegating more of the tile binding and backend lowering to the compiler.

MLIR-AIR~\cite{wang2025mlir-air} provides a platform-agnostic compiler abstraction for spatial accelerators that is organized around loop structure. AIR uses \texttt{air.herd}s to represent spatial loops mapped across tile arrays, and \texttt{scf.for} loops to represent temporal iteration within each tile. Communication and synchronization are expressed explicitly through \texttt{air.channel}s and \texttt{token}s.
AIR's loop-nest structure is useful for implementing the fused kernel, because this representation makes the spatial and temporal structure of a tiled computation visible to the compiler, together with the reuse scope and lifetime of intermediate data.

The use of these programming environments reflects the provenance of the evaluated FlashAttention implementations rather than an intended comparison of compiler abstractions.
Specifically, we evaluate the original DATO implementation, the public IRON MHA/GQA reference design\footnotemark[3]\footnotetext[3]{Public IRON MHA/GQA reference design. GitHub: \textit{blocked for double-blind review}.}, and the public MLIR-AIR kernel-fusion FlashAttention reference design\footnotemark[4]\footnotetext[4]{Public MLIR-AIR kernel-fusion FlashAttention reference design. GitHub: \textit{blocked for double-blind review}.}. These implementations serve as concrete points in the mapping progression. The three environments expose different balances between direct hardware control and compiler-managed orchestration. IRON gives programmers explicit control over tile placement, buffering, DMA operations, and runtime sequencing. DATO expresses task-and-stream dataflow while delegating more placement and orchestration to compiler lowering. MLIR-AIR keeps spatial and temporal structure explicit through herds, loops, channels, and dependencies while automating device-specific lowering. These are programming-workflow tradeoffs, and we do not claim that a mapping used by one implementation cannot be expressed in another.

None of these systems automatically discovers the mappings evaluated in this paper. Rather, they provide abstractions and lowering infrastructure for capturing an expert-selected mapping and reproducibly realizing its tile organization, communication, synchronization, and runtime execution.

\section{Related Work}
\label{sec:related}

Prior work on AMD AI Engine and XDNA devices spans compiler accessibility, architecture-aware kernel optimization, and workload-specific spatial pipelines.
DATO and IRON are particularly relevant to this work because they present FlashAttention implementations on AMD XDNA.~\cite{hunhoff2025iron,fang2025dato}
More broadly, prior work on computation mapping to AMD NPUs and AI Engine platforms provides useful insight into how performance is shaped by data movement and on-chip communication, but focus on GEMM, and similar single-kernels rather than on multi-stage fused workloads such as attention.~\cite{roesti2025unlocking,taka2026striking,mhatre2025gama,wang2025atb,binder2025architectureaware,brown2025seamless,nozaki2024ntt,xu2025tilelevel,du2026gemma3edge}
Brown and Rodr\'iguez Canal lower Fortran intrinsics through Flang and MLIR to AIE implementations, reducing the architecture-specific knowledge required of source programmers.~\cite{10.1145/3706628.3708854}
Binder~\emph{et al.} develop architecture-aware GEMM models that account for local and remote memory access, tiling, layouts, and collective communication across compute and memory tiles.~\cite{10.1145/3754598.3754612}
These works establish the importance of compiler support and memory-aware scheduling, but focus primarily on predefined intrinsic offload or single-kernel GEMM optimization.

Beyond GEMM, Xu~\emph{et al.} propose a tile-level pipeline for stencil computation that reduces external-memory pressure and scales across AIE tiles, while Nozaki~\emph{et al.} study NTT mapping under irregular and changing memory-access patterns.~\cite{10.1145/3706628.3708822, 10819592}
These studies demonstrate that spatial placement, streaming, and distributed-memory organization are important across scientific and cryptographic workloads.
Recent concurrent work also maps Gemma3 end to end onto Ryzen AI NPUs using hardware-aware matrix multiplication, pipelined attention, fusion, and quantization techniques.~\cite{du2026mappinggemma3edgedataflow}

More broadly, prior spatial-accelerator research has studied programming abstractions, design mappings onto distributed resources, and fusion-dataflow analysis. Our work builds on these established principles rather than claiming them individually as novel. Its contribution is a practical account of using MLIR-AIR and IRON to realize them together for a multi-stage attention design, including explicit DMA orchestration, distributed intermediate placement, conflicting broadcast patterns, and cascade-based reduction across XDNA generations.

\section{Attention Mapping Strategies on XDNA}
\label{sec:strategies}

\begin{figure*}[t]
\centering
\input{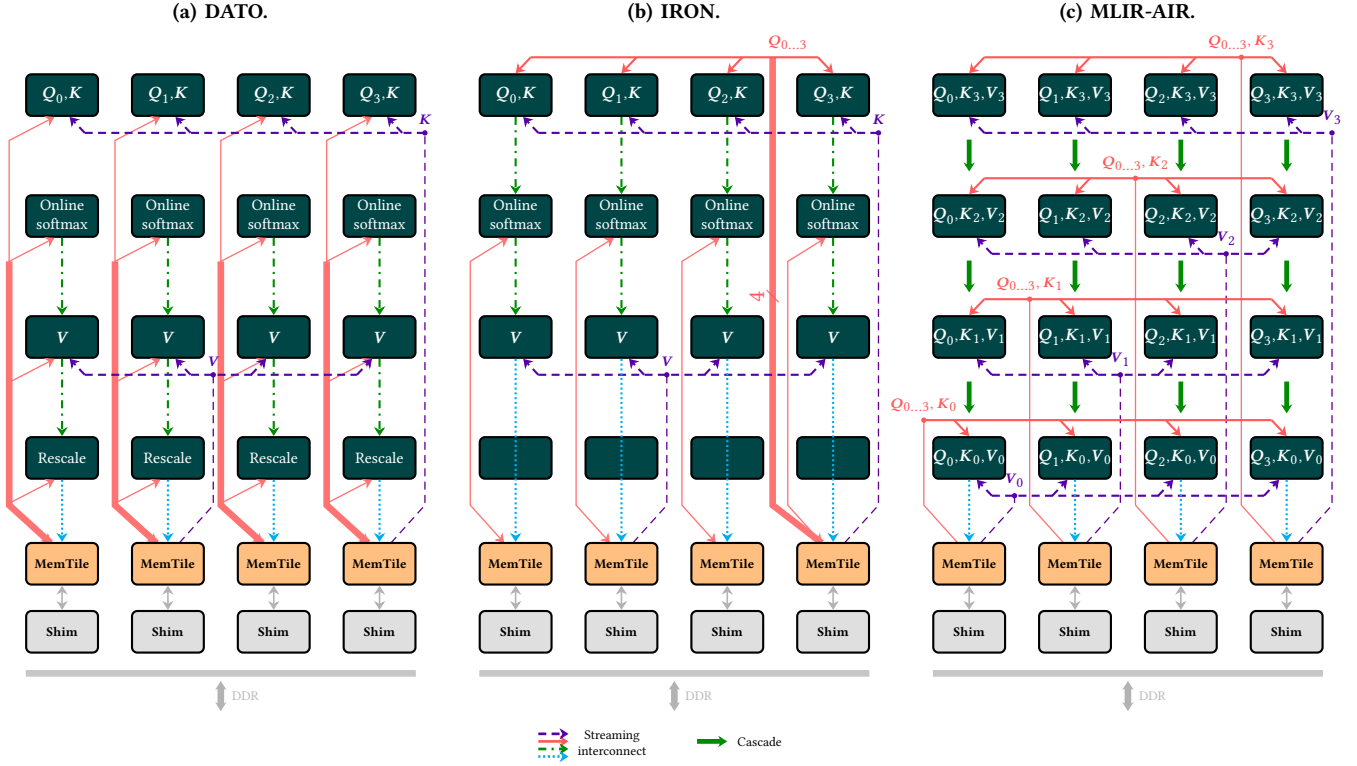}
\caption{Data path mappings for (a) DATO, (b) IRON, and (c) MLIR-AIR on a 4$\times$4 AMD XDNA tile array.}
\label{fig:dataflow}
\end{figure*}

Given the candidate mapping dimensions $H$, $l_q$, $l_k$, $d_v$, and $d_k$ discussed in Section~\ref{sec:flashattn}, we examine one na\"{\i}ve attention baseline and three FlashAttention mappings with increasing cross-stage integration. The baseline represents the simplest bring-up path, composing independently implemented operators. DATO, IRON, and MLIR-AIR progressively refine this mapping by introducing tiling over selected dimensions to improve tile-local reuse and reduce data movement.

In principle, all five dimensions may be tiled across an array of spatial accelerators. In practice, however, $H$, $d_k$, and $d_v$ are often limited in size, particularly for small and medium-sized LLMs, and therefore offer limited opportunity for on-chip reuse when distributed across a tile array. By contrast, $l_q$ and $l_k$ are typically much larger, especially at long sequence lengths, and thus form the primary dimensions of interest for spatial scaling. Accordingly, the strategies developed in this section focus on decompositions over $l_q$ and $l_k$, with $l_k$ being particularly important because it is the dominant outer reduction dimension in FlashAttention.

\subsection{Diagnostic Baseline: Unfused Layer-by-Layer Attention}
\label{sec:layer-by-layer}

The baseline strategy maps attention as four distinct operators which the implementation dispatches sequentially and executes independently of one another: the score GEMM, causal masking, the softmax, and the value-projection GEMM.
Each operator occupies the full NPU and independently tiles its inputs, thereby maximizing device utilization within a single layer but not across layers.
Intermediate results pass between operators via DDR.
This approach reconfigures the device between layers to reload the program and dataflow configuration for the next operator.
For larger sequence lengths, each operator may be dispatched multiple times to handle subslices of the larger input.

This non-FlashAttention mapping is straightforward to implement and provides a useful reference point.
However, on-chip data reuse is low because intermediate tensors are spilled to DDR between operators.
Per-stage reconfiguration further introduces extra overhead.
As a result, this strategy incurs substantial communication cost and control overhead relative to the more integrated mappings introduced here.

\subsection{Streamed Dataflow (DATO)}
\label{sec:dato}

DATO maps attention by partitioning the query dimension $l_q$ across columns of the NPU array, while tiling the $\boldsymbol{K}/\boldsymbol{V}$ sequence dimension $l_k$ temporally. As shown in Figure~\ref{fig:dato}, each column implements a staged pipeline in which four cores are assigned to the score GEMM, online softmax, value-projection GEMM, and rescaling, respectively. These stages communicate through MemTile memory, and MemTile DMA engines perform the data layout transformations required by the vectorization patterns of each stage.

Functionally, the computation performed within each column is equivalent to Algorithm~\ref{alg:cascade-stage}. Relative to Algorithm~\ref{alg:cascade-merge}, this mapping corresponds to the special case $C=1$, i.e., temporal tiling over $l_k$ without an additional spatial reduction across context partitions. Compared with the baseline, this strategy improves tile-local reuse and reduces off-chip traffic by retaining intermediate state on MemTile memory.

Its main limitation is that intermediate tensors are still exchanged between tasks through MemTile memory rather than retained within compute-tile local memory. As a result, although the data remains on chip, it continues to consume inter-tile bandwidth as it moves between pipeline stages.

\subsection{Streamed Dataflow (IRON)}
\label{sec:iron-pipeline}

IRON tiles FlashAttention along $l_q$ and $l_k$ in a similar scheme as DATO, as shown in Figure~\ref{fig:iron}. A key difference is that the rescaling operation is fused into the value-projection GEMM stage, allowing it to share compute-tile local storage across the two GEMM stages.
Inter-stage communication remains on chip but still traverses the MemTiles at each $\boldsymbol{K}/\boldsymbol{V}$ chunk iteration.

In the evaluated IRON reference design, stages are statically assigned to a three-core pipeline, and communication, buffering, and layout transformation are explicitly organized through DMA-managed transfers.
It also reduces MemTile traffic by fusing rescaling with the value-projection GEMM, so fewer intermediates are exchanged through MemTile memory.
However, intermediate score tiles are still exchanged between stages through MemTile memory rather than retained entirely within compute-tile memory. Consequently, execution remains sensitive to on-chip communication bandwidth, since $\boldsymbol{QK}^{\mathsf T}$ intermediates must be buffered, transferred, and reformatted at every $l_k$ tile.

\subsection{Fused Kernel (MLIR-AIR)}
\label{sec:fused-kernel}

Unlike the prior staged mappings, the fused-kernel strategy maps all three stages---score GEMM, online softmax, and value-projection GEMM---onto every compute tile, so that intermediate $\boldsymbol{QK}^{\mathsf T}$ scores remain in compute-tile memory throughout execution.

As shown in Figure~\ref{fig:fused}, this mapping realizes the full hierarchical FlashAttention decomposition introduced in Section~\ref{sec:flashattn}: $l_q$ is tiled spatially across columns, while $l_k$ is partitioned across tiles and reduced through tile-to-tile cascade connection: a physical interconnect that the evaluated staged mappings do not use for this reduction due to spilling intermediate data through MemTile memory.

We organize the computation around a $4 \times 4$ herd of 16 compute tiles per attention-head group. Within each herd, each row is assigned a distinct query block, while each column is assigned a disjoint subset of $\boldsymbol{K}/\boldsymbol{V}$ chunks. Consequently, tiles in the same column reuse the same $\boldsymbol{Q}$ block, whereas tiles in the same row reuse the same $\boldsymbol{K}$ and $\boldsymbol{V}$ blocks.
On XDNA~2, the full $4 \times 8$ array is populated by two $4 \times 4$ herds, allowing two attention heads to be processed in parallel.

Each tile executes the per-tile online FlashAttention kernel equivalent to Algorithm~\ref{alg:cascade-stage}. Partial results produced by different tiles are then merged across cascade-connected tiles using the tile-to-tile online-softmax reduction of Algorithm~\ref{alg:cascade-merge}.

Once configured, the fused mapping operates as a self-coordinating two-dimensional push network. DMA engines and stream routes deliver operands to the compute tiles, while tile programs consume them, maintain the online-softmax state locally, and propagate partial results through cascade links. Performance therefore depends on coordinating computation, buffering, routing, and synchronization as a single spatial schedule.

Realizing this network requires careful operand delivery.
Each compute tile has only two inbound DMA channels, yet the fused kernel must supply $\boldsymbol{Q}$, $\boldsymbol{K}$, and $\boldsymbol{V}$.
Under the $l_q \times l_k$ 2D mapping, $\boldsymbol{Q}$ is reused across columns, whereas $\boldsymbol{K}$ and $\boldsymbol{V}$ are reused across rows. We place $\boldsymbol{Q}$ and $\boldsymbol{K}$ on one shared inbound data path and reserve the second path for $\boldsymbol{V}$. Both paths broadcast from one source tile to four destinations.
While the shared $\boldsymbol{Q} / \boldsymbol{K}$ path leads to mismatch on $\boldsymbol{Q}$'s preferred broadcast directions, we duplicate the $\boldsymbol{Q}$ broadcast once per column. Each tile then selects the query block corresponding to its assigned column.
Despite increased $\boldsymbol{Q}$ traffic, the overhead is amortized because $\boldsymbol{Q}$ is loaded once per query-block group, whereas $\boldsymbol{K}$ and $\boldsymbol{V}$ dominate the streamed traffic over big sequence lengths.

Because $\boldsymbol{QK}^{\mathsf T}$ scores remain in compute-tile memory in the GEMM-native tiled layout, softmax operates directly on that layout without an intermediate re-tiling step.
This enables partial results being merged through hardware cascade connections rather than MemTile DMA. As a result, only $\boldsymbol{Q}$, $\boldsymbol{K}$, and $\boldsymbol{V}$ are streamed into compute tiles, and only the final output is written back. Compared with DATO and IRON, this removes the repeated MemTile memory traffic associated with inter-stage score exchange and layout transformation. The resulting mapping therefore improves local data reuse, and avoids consuming MemTile bandwidth for intermediate reductions.

For GQA, DDR traffic scales with the number of distinct $K/V$ heads, $H_{kv}$, rather than the number of query heads, $H$. Within the NPU, the compiler replicates each shared $K/V$ head across the spatial query-head partitions that consume it. This deliberate placement uses the distributed on-chip memories to keep data near the compute and preserve parallel execution.

\begin{table}[t]
\caption{Bandwidth ceilings at each memory hierarchy level.}
\label{tab:bw-ceilings}
\centering
\small
\begin{tabular}{llcSS}
\toprule
\textbf{Arch.} & \makecell{\textbf{DMA}\\\textbf{count}} & \makecell{\textbf{Channel}\\\textbf{count}} & {\makecell{\textbf{Bandwidth}\\\textbf{(GB/s)}}} & {\makecell{\textbf{Ridge}\\\textbf{(FLOP/B)}}} \\
\midrule
\multirow{3}{*}{XDNA~1} & $4\times$ Shim tile & 4 $\times$ 2 & 32.0 & 32.1 \\
 & $4\times$ MemTile & 4 $\times$ 6 & 96.0 & 10.7 \\
 & $16\times$ Compute tile & 16 $\times$ 2 & 128.0 & 8.0 \\
\midrule
\multirow{3}{*}{XDNA~2} & $8\times$ Shim tile & 8 $\times$ 2 & 83.2 & 101.0 \\
 & $8\times$ MemTile & 8 $\times$ 6 & 249.6 & 33.7 \\
 & $32\times$ Compute tile & 32 $\times$ 2 & 332.8 & 25.2 \\
\bottomrule
\end{tabular}
\end{table}

\section{Roofline-Driven Analysis}
\label{sec:roofline}

The four mappings considered in Section~\ref{sec:strategies} differ substantially in how the intermediates are stored and communicated: spilled to DDR, spilled to MemTile memory, or staged in compute-tile memory. Their performance implications can be explained using the roofline model, with throughput on y-axis and operational intensity (OI) on x-axis, in FLOP/B. At low OI, performance lies on the bandwidth-bound segment. Beyond the ridge point, performance enters the compute-bound segment, where throughput is limited by the compute ceiling. Increasing OI therefore drives the mapping rightward on the roofline and raises attainable throughput until the compute ceiling is reached.

\subsection{The Operational Intensity Constraint}
\label{sec:roofline_qk}

In attention, intermediate data movement can significantly constrain OI.
The dominant workload lies in the two GEMMs, with workloads $2 H l_q l_k d_k$ and $2 H l_q l_k d_v$, respectively.
The $\boldsymbol{QK}^{\mathsf T}$ GEMM's write-side OI is $d_k$ in \texttt{bfloat16}, meaning that the write-side OI is bounded by $d_k$~FLOP/B if spilling $\boldsymbol{QK}^{\mathsf T}$ alone. Spilling any additional intermediates tightens this constraint further.
Reducing intermediate traffic is therefore the key to optimizing the throughput performance.

\begin{table*}[t]
\caption{Asymptotic OI bounds and memory traffic at each NPU memory hierarchy. Tensors are in \texttt{bfloat16}.}
\label{tab:oi-values}
\centering
\small
\begin{tabular}{c cc cc cc}
\toprule
 & \multicolumn{2}{c}{\textbf{Shim tile}} & \multicolumn{2}{c}{\textbf{MemTile}} & \multicolumn{2}{c}{\textbf{Compute tile}} \\
\cmidrule(lr){2-3} \cmidrule(lr){4-5} \cmidrule(lr){6-7}
\textbf{Strategy} & \textbf{Matrices spilled} & \textbf{FLOP/B} & \textbf{Matrices spilled} & \textbf{FLOP/B} & \textbf{Matrices spilled} & \textbf{FLOP/B} \\
\midrule
Layer by layer & $\boldsymbol{Q}, \boldsymbol{K}, \boldsymbol{V}, \boldsymbol{QK}^{\mathsf T}, \boldsymbol{G}$, scale & $< \nicefrac{d_k}{2}$ & $\boldsymbol{Q}, \boldsymbol{K}, \boldsymbol{V}, \boldsymbol{QK}^{\mathsf T}, \boldsymbol{G}$, scale & $< \nicefrac{d_k}{2}$ & $\boldsymbol{Q}, \boldsymbol{K}, \boldsymbol{V}, \boldsymbol{QK}^{\mathsf T}, \boldsymbol{G}$, scale & $< \nicefrac{d_k}{2}$ \\
DATO & $\boldsymbol{Q}, \boldsymbol{K}, \boldsymbol{V}$ & $\propto$ Seq. Len. & $\boldsymbol{Q}, \boldsymbol{K}, \boldsymbol{V}, \boldsymbol{QK}^{\mathsf T}, \boldsymbol{G}$, scale & $< \nicefrac{d_k}{2}$ & $\boldsymbol{Q}, \boldsymbol{K}, \boldsymbol{V}, \boldsymbol{QK}^{\mathsf T}, \boldsymbol{G}$, scale & $< \nicefrac{d_k}{2}$ \\
IRON & $\boldsymbol{Q}, \boldsymbol{K}, \boldsymbol{V}$ & $\propto$ Seq. Len. & $\boldsymbol{Q}, \boldsymbol{K}, \boldsymbol{V}, \boldsymbol{QK}^{\mathsf T}, \boldsymbol{G}$ & $\leq \nicefrac{d_k}{2}$ & $\boldsymbol{Q}, \boldsymbol{K}, \boldsymbol{V}, \boldsymbol{QK}^{\mathsf T}, \boldsymbol{G}$ & $\leq \nicefrac{d_k}{2}$ \\
MLIR-AIR & $\boldsymbol{Q}, \boldsymbol{K}, \boldsymbol{V}$ & $\propto$ Seq. Len. & $\boldsymbol{Q}, \boldsymbol{K}, \boldsymbol{V}$ & $\leq d_k$ & $\boldsymbol{Q}, \boldsymbol{K}, \boldsymbol{V}$ & $\leq d_k$ \\
\bottomrule
\end{tabular}
\end{table*}

\begin{table*}[t]
\caption{Predicted execution regime at $d_k=64$, derived by comparing the OI bounds in Table~\ref{tab:oi-values} with the ridge points in Table~\ref{tab:bw-ceilings}. The prediction assumes the long-sequence regime, $\mathrm{Seq.\ Len.} \gg d_k$. \textbf{MB} denotes memory-bound, \textbf{CB} denotes compute-bound, and \textbf{NR} denotes near ridge point. \textbf{MB} and \textbf{NR} are highlighted because they are the most likely performance-limiting regimes in practice.}
\label{tab:oi-prediction}
\centering
\small
\begin{tabular}{c c cSc cSc cSc}
\toprule
\multirow{3}{*}{\textbf{Arch.}} & \multirow{3}{*}{\textbf{Strategy}}
& \multicolumn{3}{c}{\textbf{Shim tile}}
& \multicolumn{3}{c}{\textbf{MemTile}}
& \multicolumn{3}{c}{\textbf{Compute tile}} \\
\cmidrule(lr){3-5} \cmidrule(lr){6-8} \cmidrule(lr){9-11}
&
& \makecell{\textbf{Pred. OI}\\\textbf{(FLOP/B)}} & {\makecell{\textbf{Ridge}\\\textbf{(FLOP/B)}}} & \makecell{\textbf{Pred.}\\\textbf{Regime}}
& \makecell{\textbf{Pred. OI}\\\textbf{(FLOP/B)}} & {\makecell{\textbf{Ridge}\\\textbf{(FLOP/B)}}} & \makecell{\textbf{Pred.}\\\textbf{Regime}}
& \makecell{\textbf{Pred. OI}\\\textbf{(FLOP/B)}} & {\makecell{\textbf{Ridge}\\\textbf{(FLOP/B)}}} & \makecell{\textbf{Pred.}\\\textbf{Regime}} \\
\midrule
\multirow{4}{*}{XDNA~1}
& Layer by layer & $< 32$ & 32.1 & \textbf{NR}
                  & $< 32$ & 10.7 & CB
                  & $< 32$ & 8.0  & CB \\
& DATO           & $\propto$ Seq. Len. & 32.1 & CB
                  & $< 32$ & 10.7 & CB
                  & $< 32$ & 8.0  & CB \\
& IRON           & $\propto$ Seq. Len. & 32.1 & CB
                  & $\le 32$ & 10.7 & CB
                  & $\le 32$ & 8.0  & CB \\
& MLIR-AIR          & $\propto$ Seq. Len. & 32.1 & CB
                  & $\le 64$ & 10.7 & CB
                  & $\le 64$ & 8.0  & CB \\
\midrule
\multirow{3}{*}{XDNA~2}
& Layer by layer & $< 32$ & 101.0 & \textbf{MB}
                  & $< 32$ & 33.7 & \textbf{NR}
                  & $< 32$ & 25.2 & \textbf{NR} \\
& IRON           & $\propto$ Seq. Len. & 101.0 & CB
                  & $\le 32$ & 33.7 & \textbf{NR}
                  & $\le 32$ & 25.2 & \textbf{NR} \\
& MLIR-AIR          & $\propto$ Seq. Len. & 101.0 & CB
                  & $\le 64$ & 33.7 & CB
                  & $\le 64$ & 25.2 & CB \\
\bottomrule
\end{tabular}
\end{table*}

\subsection{Roofline Model at Three Memory Levels}
\label{sec:three-level-roofline}

The AMD XDNA NPU exposes multiple bandwidth domains, and a single roofline is therefore insufficient to characterize attention mappings. In particular, data movement may be limited by shim-tile, MemTile or compute-tile DMA bandwidths. These three levels differ substantially in both bandwidth and in the role they play in the mappings of Section~\ref{sec:strategies}.

We therefore analyze the four strategies with respect to three separate rooflines: shim-tile, MemTile, and compute-tile DMA bandwidths. This separation is necessary because a mapping may relieve pressure at one level while remaining constrained at another. For example, moving intermediate tensors from DDR to MemTile memory improves the shim-tile roofline position but may still leave the mapping limited by MemTile DMA traffic. Similarly, a fused kernel can eliminate MemTile communication for intermediates yet remain constrained by operand delivery into compute tiles. Examining the three memory levels separately thus makes it possible to identify which hierarchy level is performance-limiting for each strategy, and if any further on-chip reuse needs to be pursued to improve OI at that level.

\paragraph{Compute ceiling.}
In this paper, we derive the compute ceiling from instruction counting on the fused kernel rather than from the theoretical hardware peak. On XDNA~1, each $64{\times}64$ $\boldsymbol{K/V}$ tile performs approximately $1.06 \times 10^6$ FLOPs but requires about 16504 scalar and vector instructions, yielding 64.2 FLOP/cycle/tile and an overall ceiling of 1.03~TFLOP/s at 1.0~GHz across 16 tiles. On XDNA~2, the same tile requires 5304 instructions and performs approximately $1.07 \times 10^6$ FLOPs, yielding 202 FLOP/cycle/tile and an overall ceiling of 8.4~TFLOP/s at 1.3~GHz across 32 tiles.

This compute ceiling is difficult to approach in practice. On XDNA~2, for example, each tile of Algorithm~\ref{alg:cascade-stage} requires 5304 instructions, of which approximately 3000 are softmax operations. These instructions contribute relatively few FLOPs compared with the much denser GEMM operations. In addition, the irregular arithmetic and row-wise dependences in softmax make it difficult to sustain one issued instruction per cycle. As a result, inefficiencies in the softmax stage have a disproportionate effect on overall throughput.

\paragraph{Bandwidth ceilings.}
Each DMA channel transfers up to 32~bits/cycle. Table~\ref{tab:bw-ceilings} derives the aggregate bandwidth at each memory level from the hardware channel counts.

\subsection{Predicted Performance}
\label{sec:predicted-vs-measured}

Table~\ref{tab:oi-values} summarizes the asymptotic OI bounds of the four strategies at each NPU memory level, derived from the tensors spilled by each mapping.
Table~\ref{tab:oi-prediction} then evaluates these bounds at $d_k=64$ and compares them against the roofline ridge points, yielding a theoretical prediction of whether each mapping is compute-bound, memory-bound, or near the ridge at each level.

The predicted regimes differ across architectures.
On XDNA~1, the ridge points are low enough that most mappings are compute-bound. On XDNA~2, IRON remains constrained by MemTile bandwidth, and only the fused mapping clearly reaches the compute-bound regime.
The next section compares these predictions against measured performance.

\section{Evaluation}
\label{sec:eval}

In this section, we present the measured results of the four attention mappings, and examine whether the measured trends are consistent with the OI shifts and memory-level bottlenecks predicted by the roofline models presented in the previous section.

\subsection{Experimental Setup}
\label{sec:eval_setup}

We evaluate the four attention mappings on two AMD XDNA NPU generations: XDNA~1 (Ryzen 7 8845HS) and XDNA~2 (Ryzen AI 9 HX 370). We report results from the GPT-2 Small attention dimensions~\cite{radford2019language}: 12 heads, $l_q = l_k =$ sequence length, $d_k = d_v = 64$, and causal masking enabled. Data movement and computation use \texttt{bfloat16}.
On XDNA~2, each mapping shown in Figure~\ref{fig:dataflow} is scaled to occupy all eight device columns.

\begin{table*}[t]
\caption{GPT-2 Small attention performance of the four mapping strategies on XDNA~1 and XDNA~2. DATO only reports XDNA~1 results.}
\label{tab:perf-causal-combined}
\centering
\footnotesize
\begin{tabular}{c S[table-format=5.0] S[table-format=4.1] S[table-format=3.0] S[table-format=4.1] S[table-format=2.0] S[table-format=4.1] S[table-format=4.0] S[table-format=3.1] S[table-format=4.0] c}
\toprule
\multirow{2}{*}{\textbf{Platform}} & {\multirow{2}{*}{\textbf{Seq. Len.}}}
& \multicolumn{2}{c}{\textbf{Layer by layer}}
& \multicolumn{2}{c}{\textbf{DATO}}
& \multicolumn{2}{c}{\textbf{IRON}}
& \multicolumn{3}{c}{\textbf{MLIR-AIR}} \\
\cmidrule(lr){3-4} \cmidrule(lr){5-6} \cmidrule(lr){7-8} \cmidrule(lr){9-11}
& & {\makecell{\textbf{Latency}\\\textbf{(ms)}}} & \textbf{GFLOP\textsubscript{GEMM}/s}
  & {\makecell{\textbf{Latency}\\\textbf{(ms)}}} & \textbf{GFLOP\textsubscript{GEMM}/s}
  & {\makecell{\textbf{Latency}\\\textbf{(ms)}}} & \textbf{GFLOP\textsubscript{GEMM}/s}
  & {\makecell{\textbf{Latency}\\\textbf{(ms)}}} & \textbf{GFLOP\textsubscript{GEMM}/s} & \textbf{$\Delta$ vs. IRON} \\
\midrule
\multirow{5}{*}{\textbf{XDNA~1}}
& 512   &    22.0 &   37 &   200.2 & 4   &    9.8 & 82  &    7.0 & 114 & \bfseries 1.39$\times$ \\
& 1024  &    42.1 &   77 &   225.0 & 14  &   30.8 & 105 &   20.1 & 160 & \bfseries 1.53$\times$ \\
& 2048  &   112.8 &  114 &   499.7 & 26  &   83.4 & 154 &   67.7 & 190 & \bfseries 1.23$\times$ \\
& 4096  &   378.0 &  136 &  1220.1 & 42  &  264.6 & 195 &  250.0 & 206 & \bfseries 1.06$\times$ \\
& 8192  &  3073.8 &   67 &  4854.1 & 42  & 1054.0 & 196 &  955.2 & 216 & \bfseries 1.10$\times$ \\ %
\midrule
\multirow{8}{*}{\textbf{XDNA~2}}
& 256    &    2.0 & 101 & {---} & {---} &    1.1 & 177  &    0.3 & 629  & \bfseries 3.91$\times$ \\
& 512    &   10.1 & 222 & {---} & {---} &    1.4 & 596  &    0.6 & 1241 & \bfseries 2.08$\times$ \\
& 1024   &   15.4 & 347 & {---} & {---} &    3.4 & 947  &    1.7 & 1844 & \bfseries 1.95$\times$ \\
& 2048   &   35.8 & 417 & {---} & {---} &   10.1 & 1277 &    5.2 & 2486 & \bfseries 1.95$\times$ \\
& 4096   &  112.2 & 473 & {---} & {---} &   34.1 & 1513 &   17.2 & 2997 & \bfseries 1.98$\times$ \\
& 8192   &  440.8 & 469 & {---} & {---} &  123.7 & 1668 &   62.0 & 3324 & \bfseries 1.99$\times$ \\
& 16384  & 1780.0 & 463 & {---} & {---} &  469.5 & 1757 &  234.5 & 3517 & \bfseries 2.00$\times$ \\
& 32768  & 7251.7 & 455 & {---} & {---} & 1828.1 & 1805 &  911.1 & 3622 & \bfseries 2.01$\times$ \\
\bottomrule
\end{tabular}
\end{table*}

\begin{figure*}[h!]
\centering
\includegraphics[width=\textwidth]{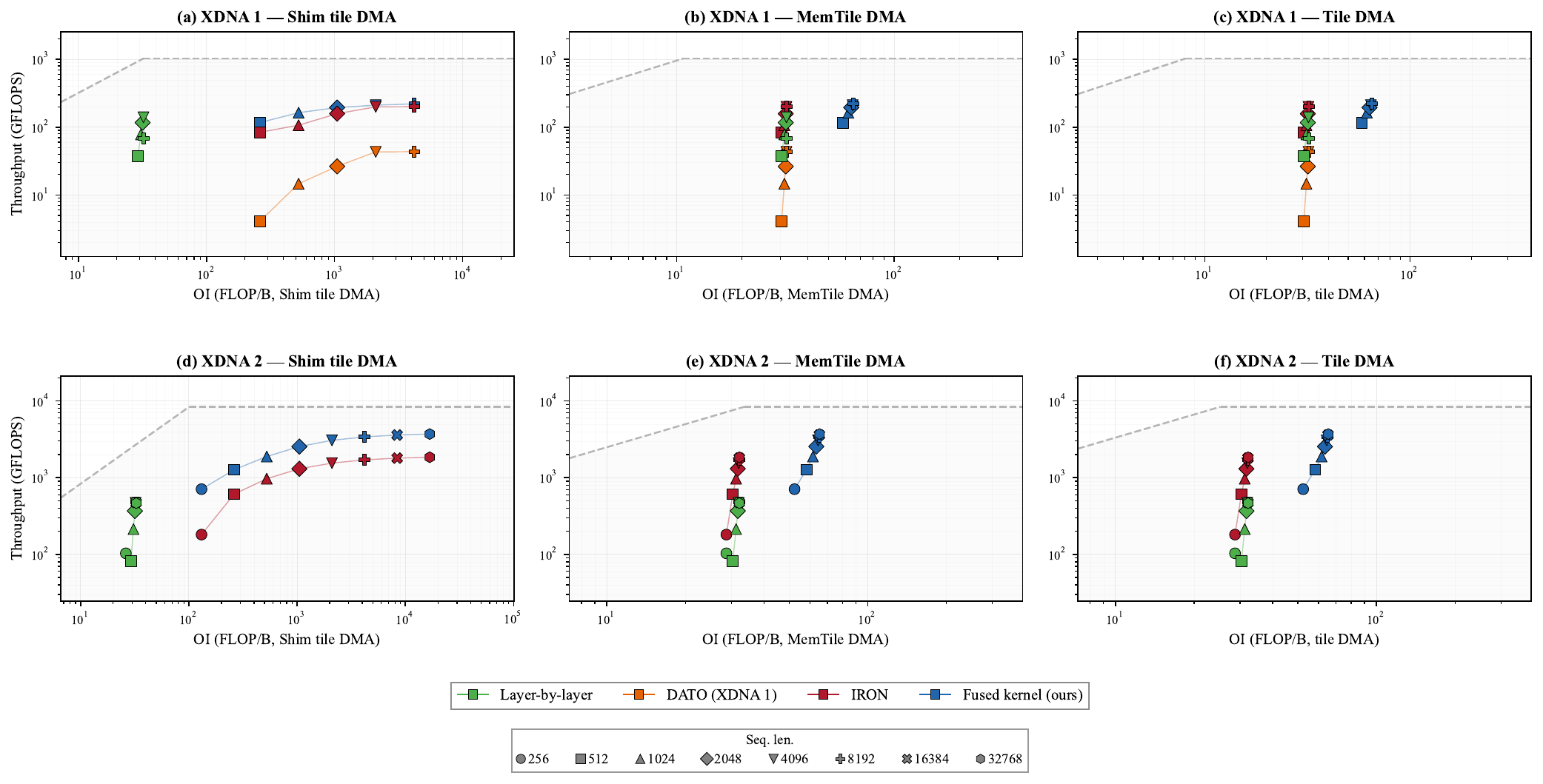}
\caption{Roofline analysis of the four mapping strategies at three memory hierarchy levels on XDNA~1 and 2.}
\label{fig:roofline}
\end{figure*}

For each mapping, we measure execution time over a range of sequence lengths and convert the results to effective throughput.
We report effective GEMM throughput,
\begin{equation}
    P_{effGEMM} = \frac{2HL_qL_k(d_k+d_v)}{T_{attention}}
\end{equation}
where $T_{attention}$ is the end-to-end FlashAttention latency. The timed region includes both GEMMs, online softmax, and runtime dispatch overheads.

We use GEMM work in the numerator because it measures the sustained computation by the vectorized matrix-engine, while excluding the softmax computation which lacks a standard FLOP-counting convention.
Each configuration is evaluated over 20 iterations after 10 warmup iterations, and we report the minimum latency to reduce the effect of operating-system and DDR-induced runtime variation.

\subsection{Measured Performance}
\label{sec:eval_perf}

Tables~\ref{tab:perf-causal-combined} report the measured latency and effective GEMM throughput of the four mappings across sequence lengths. The results follow the same overall ordering on both NPU generations: the layer-by-layer baseline attains the lowest throughput, staged on-chip mappings (DATO and IRON) improve upon the baseline, and the MLIR-AIR fused kernel attains the highest throughput. This trend is consistent with the progression in Section~\ref{sec:strategies}, where increasingly optimized mappings retain more intermediate data on chip and reduce communication through levels of the memory hierarchy.

The four-mapping comparison in Tables~\ref{tab:perf-causal-combined} is limited to $d_k=64$. The relative ordering reported here is therefore established only for that configuration.
We separately evaluate the fused MLIR-AIR implementation at $d_k=128$. Both MHA and GQA remain compute-bound at the shim, MemTile, and compute-tile levels. However, corresponding $d_k=128$ implementations for the other mappings are unavailable.

Overall, the measured results support the qualitative claim of the previous sections: tighter on-chip retention of intermediates leads to higher throughput. The next subsection examines whether these measured trends are consistent with the memory-level bottlenecks predicted by the roofline model.

\subsection{Validation Against the Roofline Model}
\label{sec:eval_roofline}

Figure~\ref{fig:roofline} projects the measured performance of the four mappings onto the shim-tile, MemTile, and compute-tile rooflines on XDNA~1 and XDNA~2.
Beyond confirming the overall trend, these plots provide a first-principles interpretation of the measurements.
Specifically, they show how each mapping changes OI at each memory level and, in turn, how that change shifts the attainable peak performance under the roofline model.

For consistency with the roofline model, throughputs in Figure~\ref{fig:roofline} are recalculated using the full attention workload rather than the GEMM-only workload reported in Table~\ref{tab:perf-causal-combined}. In particular, the projected throughput includes the softmax computation and uses the same workload definition as that used to derive the compute ceilings in Section~\ref{sec:three-level-roofline}.

Figures~\ref{fig:roofline}a---c show the rooflines on XDNA~1.
In the shim-tile plot of Figure~\ref{fig:roofline}a, the DATO, IRON, and MLIR-AIR mappings shift rightward with sequence length and move toward the compute-bound region, consistent with the prediction in Table~\ref{tab:oi-values} that their shim-tile OI increases with sequence length.
By contrast, the layer-by-layer baseline remains at substantially lower OI, close to the ridge point, and its throughput is correspondingly capped at lower values.
In the MemTile and compute-tile rooflines shown in Figures~\ref{fig:roofline}b--c, all mappings lie in the compute-bound regime.
This is consistent with the prediction in Table~\ref{tab:oi-prediction}.

Figures~\ref{fig:roofline}d--f show the rooflines on XDNA~2.
The measured results again follow the predictions in Table~\ref{tab:oi-prediction}, but with a different bottleneck structure from XDNA~1.
In Figure~\ref{fig:roofline}d, the layer-by-layer baseline remains well within the memory-bound regime, and its throughput is correspondingly limited by shim-tile DMA bandwidth.
The MemTile roofline in Figure~\ref{fig:roofline}e shows that IRON remains clustered near the ridge point, indicating MemTile bandwidth-limited behavior, whereas the fused mapping shifts further right and upward into the compute-bound region.
The compute-tile roofline in Figure~\ref{fig:roofline}f shows the same separation: once the dominant intermediate traffic is removed from MemTile, the fused mapping moves further into the compute-bound regime.

Taken together, the empirical roofline plots validate the regime predictions derived from Tables~\ref{tab:oi-values} and~\ref{tab:oi-prediction}. On XDNA~1, retaining intermediates at MemTile memory is sufficient to place the on-chip mappings in the compute-bound regime. On XDNA~2, the higher ridge points make MemTile bandwidth a visible constraint for mappings that continue to exchange intermediates through L2, and only the MLIR-AIR fused kernel, which retains these intermediates at compute-tile memory, clearly moves beyond this constraint.

\subsection{Energy Efficiency}
\label{sec:eval_energy}

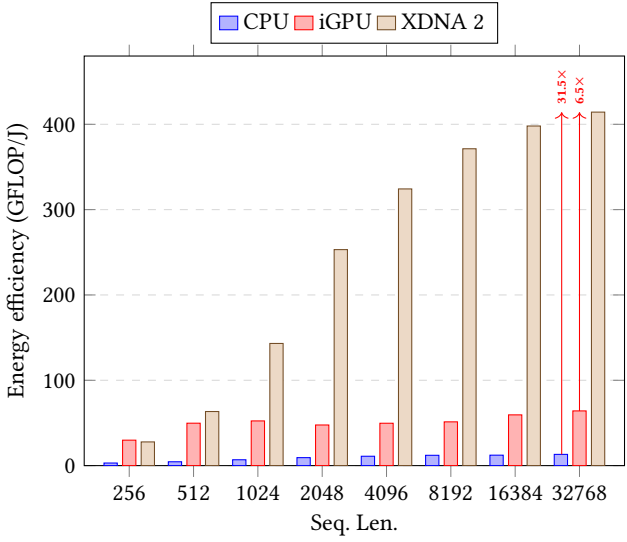
\begin{figure}[t]
\centering
\begin{tikzpicture}
\begin{axis}[
    ybar,
    bar width=5pt,
    width=1.03\columnwidth,
    height=7cm,
    ymin=0,
    ymax=480,
    enlarge x limits=0.1,
    ylabel={Energy efficiency (GFLOP/J)},
    ylabel style={yshift=-10pt},
    xlabel={Seq. Len.},
    symbolic x coords={256,512,1024,2048,4096,8192,16384,32768},
    xtick=data,
    legend style={at={(0.5,1.03)}, anchor=south, legend columns=3},
    legend image code/.code={\draw[#1] (0cm,-0.09cm) rectangle (0.24cm,0.09cm);},
    ymajorgrids=true,
    grid style={dashed,gray!30},
]
\addplot table[x=N,y=CPU,col sep=space] {data/energy_efficiencies.txt};
\addplot table[x=N,y=iGPU,col sep=space] {data/energy_efficiencies.txt};
\addplot table[x=N,y=NPU2,col sep=space] {data/energy_efficiencies.txt};

\legend{CPU,iGPU,XDNA~2}

\draw[->, thin, xshift=-6.7pt, color=red] (axis cs:32768,13.1676804) -- (axis cs:32768,414.2724943)
    node[pos=1, right, font=\tiny, rotate=90] {\textbf{31.5$\times$}};
\draw[->, thin, color=red] (axis cs:32768,64.11146518) -- (axis cs:32768,414.2724943)
    node[pos=1, right, font=\tiny, rotate=90] {\textbf{6.5$\times$}};
\end{axis}
\end{tikzpicture}
\caption{Energy efficiency of the MLIR-AIR fused kernel on XDNA~2 against CPU and iGPU.}
\label{fig:energy-eff}
\end{figure}

The throughput gains observed in the roofline analysis also translate to improved energy efficiency.
Figure~\ref{fig:energy-eff} compares the MLIR-AIR fused-kernel mapping on XDNA~2 against CPU and iGPU execution on the same Ryzen AI 9 HX 370 platform.
The iGPU was selected as a baseline because it is the most relevant alternative platform for LLM inference on the same Ryzen AI device.
The iGPU baseline is an optimized fused FlashAttention implementation written in Triton and compiled through the AMD GPU backend in ROCm\texttrademark~7.2 to the native GFX1150 code within the same Ryzen AI SoC. The NPU and iGPU implementations evaluate the same bfloat16 causal-attention configurations and use the same effective-matmul workload when computing throughput and energy efficiency.
Energy efficiency is reported in GFLOP/J and is computed from the measured throughput and the average device power collected using a profiler tool.
Across all evaluated sequence lengths, XDNA~2 consistently achieves high energy efficiency. At 32k sequence length, XDNA~2 achieves $31.5\times$ improved efficiency relative to the CPU, and $6.5\times$ relative to the iGPU.

These results are consistent with the preceding performance analysis. The MLIR-AIR fused mapping sustains higher throughput on the NPU, and this directly translates to higher energy efficiency. The effect is most pronounced at long sequence lengths, where the throughput gap relative to the CPU and iGPU is largest. These results improve the viability of XDNA~2 as an execution target for edge LLM operators.

\subsection{Generality Across LLM Configurations}
\label{sec:eval_generality}

The ablation study in the previous sections uses GPT-2 Small attention dimensions ($H{=}12$, $d_k{=}d_v{=}64$).
A natural question is whether the MLIR-AIR fused kernel strategy generalizes to other LLM architectures.
Table~\ref{tab:model-configs} evaluates the fused kernel across a range of publicly available LLM configurations, including encoder-only models, decoder-only models, and both MHA and GQA variants. The evaluated models span $d_{\text{model}}$ from 768 to 8192, head configurations from 12/12 to 64/8, and sequence lengths from 2048 to 128K.

Across these configurations, the fused kernel sustains effective GEMM throughput in the range of approximately 2.0--3.6~TFLOP\textsubscript{GEMM}/s on XDNA~2. For sequence length 2048, throughput remains above 2.0~TFLOP\textsubscript{GEMM}/s across GPT-2, BERT, Llama, and DeepSeek variants despite substantial variation in model width, head count, and attention type. At longer sequence lengths, the kernel continues to sustain high throughput on Qwen and related GQA configurations at sequence length 32768.

These results suggest that the proposed mapping generalizes across common attention configurations used in contemporary LLMs. In particular, it applies to both causal and non-causal masking, to both MHA and GQA, and to a broad range of head and hidden dimensions.

\begin{table*}[t]
\centering
\caption{Supported LLM configurations with MLIR-AIR fused kernel on XDNA~2.}
\label{tab:model-configs}
\begin{tabular}{lcccccS[table-format=6.0]S[table-format=8.0]c}
\toprule
\textbf{Model} & \textbf{$d_{\mathrm{model}}$} & \textbf{$H/H_{\mathrm{kv}}$} & \textbf{$d_k$} & \textbf{Type} & \textbf{Mask} & {\textbf{Seq. Len.}} & {\textbf{Latency ($\mu$s)}} & \textbf{Throughput (GFLOP/s)} \\
\midrule
BERT-Base        &   768 & 12/12 &  64 & MHA & Non-causal &   2048 & 5775 & 2231.15 \\
BERT-Large       & 1024 & 16/16 &  64 & MHA & Non-causal &   2048 & 7526 & 2282.74 \\
\midrule
GPT-2 Small / OPT-125M      &   768 & 12/12 &  64 & MHA & Causal     &   2048 & 5810 & 2217.71 \\
GPT-2 Medium     & 1024 & 16/16 &  64 & MHA & Causal     &   2048 & 7535 & 2280.01 \\
GPT-2 Large      & 1280 & 20/20 &  64 & MHA & Causal     &   2048 & 9196 & 2335.24 \\
\midrule
Llama-2 7B       & 4096 & 32/32 & 128 & MHA & Causal     &   2048 & 34201 & 2009.28 \\
Llama-3 8B       & 4096 & 32/8  & 128 & GQA & Causal     &   2048 & 34076 & 2016.65 \\
\midrule
Qwen1.5-0.5B         & 1024 & 16/16  & 64 & MHA & Causal     &  32768 & 1214700 & 3620.68 \\
Qwen2.5-0.5B-Instruct       & 896 & 14/2  & 64 & GQA & Causal     & 32768 & 1062870 & 3620.66 \\
Qwen1.5-0.5B (128K)         & 1024 & 16/16  & 64 & MHA & Causal     &  131072 & 18996700 & 3704.00 \\
\midrule
DeepSeek LLM 7B  & 4096 & 32/32 & 128 & MHA & Causal     &   4096 & 114443 & 2401.88 \\
DeepSeek LLM 67B & 8192 & 64/8  & 128 & GQA & Causal     &   4096 & 227467 & 2416.86 \\
\bottomrule
\end{tabular}
\end{table*}

The $d_k{=}64$ configurations achieve $\sim2.0$~TFLOP/s, with throughput increasing slightly as more heads amortize launch overhead. The $d_k{=}128$ configurations achieve $\sim1.8$~TFLOP/s. The larger head dimension increases per-tile memory pressure and softmax work per chunk.
Llama-3 GQA ($H_{kv}=8$) performs within 1\% of Llama-2 MHA ($H_{kv}=32$). GQA reduces the unique $K/V$ data transferred from DDR, while the compiler replicates shared K/V heads within the NPU to colocate them with the query-head compute partitions. Because the fused mapping is compute-bound, this on-chip replication preserves MHA-like throughput while retaining GQA's off-chip traffic reduction.

\subsection{Numerical Stability}
\label{sec:eval_stability}

Numerical stability must be evaluated for our FlashAttention mappings because intermediates, including $\boldsymbol{QK}^{\mathsf T}$ and the online-softmax state, are quantized in \texttt{bfloat16} for execution on the NPU vector engines.
We evaluate the numerical stability of the MLIR-AIR fused kernel by sweeping the input variance. Specifically, $\boldsymbol{Q}$, $\boldsymbol{K}$, and $\boldsymbol{V}$ are drawn independently from $\mathcal{N}(0,\sigma^2)$ for each element, with $\sigma$ varied from $0.25$ to $4.0$.
For $Q,K,V \sim \mathcal{N}(0,\sigma^2)$, each pre-softmax score
$\boldsymbol{QK}^{\mathsf T}/\sqrt{d_k}$ has variance
$\sigma^4$ under the standard attention scaling.
Increasing $\sigma$ therefore sharpens the softmax distribution and amplifies the effect of \texttt{bfloat16} truncation in the intermediate score buffer. In practice, transformer activations after layer normalization typically have $\sigma \approx 1$. Our sweep extends beyond this regime to characterize the precision boundary of the kernel.
\texttt{fp32} scaled dot-product attention with identical inputs is used as the reference.

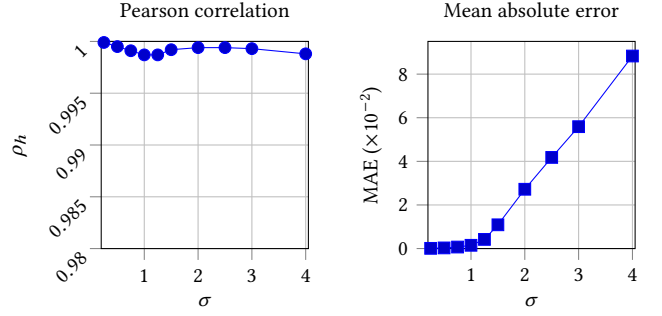
\begin{figure}[t]
\centering
\begin{tikzpicture}

\begin{axis}[
    at={(0cm,4.2cm)},
    anchor=north west,
    width=0.51\linewidth,
    height=0.51\linewidth,
    xlabel={$\sigma$},
    ylabel={$\rho_h$},
    xmin=0.2, xmax=4.05,
    ymin=0.98, ymax=1.0000,
    ytick={0.98,0.985,0.990,0.995,1.0000},
    yticklabel style={
        /pgf/number format/fixed,
        /pgf/number format/precision=4,
        rotate=45, anchor=east
    },
    grid=both,
    tick align=outside,
    tick pos=left,
    ylabel near ticks,
    enlargelimits=false,
    title={Pearson correlation},
    title style={font=\small},
    label style={font=\small},
    tick label style={font=\small},
]
\addplot+[mark=*, mark size=2.2pt] coordinates {
    (0.25,0.9999)
    (0.50,0.9995)
    (0.75,0.9991)
    (1.00,0.9987)
    (1.25,0.9987)
    (1.50,0.9992)
    (2.00,0.9994)
    (2.50,0.9994)
    (3.00,0.9993)
    (4.00,0.9988)
};
\end{axis}

\begin{axis}[
    at={(0.51\linewidth,4.2cm)},
    anchor=north west,
    width=0.51\linewidth,
    height=0.51\linewidth,
    xlabel={$\sigma$},
    ylabel={MAE ($\times 10^{-2}$)},
    xmin=0.2, xmax=4.05,
    ymin=0, ymax=9.5,
    ytick={0,2,4,6,8},
    grid=both,
    tick align=outside,
    tick pos=left,
    ylabel near ticks,
    enlargelimits=false,
    title={Mean absolute error},
    title style={font=\small},
    label style={font=\small},
    tick label style={font=\small},
]
\addplot+[mark=square*, mark size=2.2pt] coordinates {
    (0.25,0.01)
    (0.50,0.03)
    (0.75,0.07)
    (1.00,0.15)
    (1.25,0.42)
    (1.50,1.09)
    (2.00,2.72)
    (2.50,4.18)
    (3.00,5.59)
    (4.00,8.83)
};
\end{axis}

\end{tikzpicture}
\vspace{-5mm}
\caption{Numerical stability versus input standard deviation for GPT-2 Small ($d_k=64$, $H=12$, $L_Q=L_K=2048$). Inputs satisfy $Q,K,V \sim \mathcal{N}(0,\sigma^2)$. Reference: FP32 scaled dot-product attention.}
\label{fig:precision}
\end{figure}

Figure~\ref{fig:precision} reports results for the GPT-2 Small configuration ($d_k=64$, $H=12$) with $L_Q=L_K=2048$.
Mean absolute error increases from $0.0001$ at $\sigma=0.25$ to $0.0883$ at $\sigma=4.0$, consistent with the $\sigma^4$ growth in score variance. The per-head Pearson correlation $\rho_h$ remains above $0.998$ across the full sweep. These results indicate that the fused kernel remains numerically stable over the input range relevant to normalized transformer activations, with degradation appearing only under substantially larger score variance.

\section{Best Practices and Lessons Learned}
\label{sec:lessons}

The following principles are established at a high level. The practical contribution of this case study is showing how they must be realized under XDNA-specific constraints and expressed through the available open-source compiler tools. Each lesson connects a hardware constraint to the corresponding mapping decision, its implementation in the tool flow, and its measured performance consequence.

\subsection{Retain Dominant Intermediates On Chip}
\label{sec:lessons_intermediates}

A primary lesson from this study is that intermediate placement has a first-order effect on performance. When large intermediates are materialized farther from the compute tile, the resulting data movement lowers OI and can make the mapping bandwidth-limited. For attention, this effect is most pronounced for the $\boldsymbol{QK}^{\mathsf T}$ tensor, whose size scales with $l_q l_k$ and dominates the intermediate memory footprint.

From this perspective, the transition from the staged DATO or IRON mappings to the MLIR-AIR fused-kernel mapping is a fusion step. Here, fusion means combining consecutive operators into a single on-chip execution pipeline so that intermediates are consumed directly rather than spilled through the memory hierarchy. In the fused mapping, the score GEMM, online softmax, and value-projection GEMM execute within each compute tile, and the $\boldsymbol{QK}^{\mathsf T}$ intermediates remain in local memory throughout the pipeline.

\subsection{Use Multi-Level Roofline Analysis Early}
\label{sec:lessons_roofline}

A second lesson is that roofline analysis is most useful when applied \textit{(i)} before implementation, and \textit{(ii)} separately at each relevant memory level.
For a candidate mapping, applying this analysis requires only a few inputs: which tensors are spilled at each memory level, the corresponding OI, and the ridge point of that level.
This comparison gives an immediate prediction of whether the mapping is likely to be memory-bound or compute-bound, and therefore whether deeper fusion is likely to yield meaningful benefit.

Our FlashAttention case study illustrates this procedure.
On XDNA~1, the MemTile and compute-tile ridge points are low enough that staged on-chip mappings are already compute-bound once the dominant intermediates no longer spill to DDR.
In that setting, further fusion into compute-tile memory is not required to remove a bandwidth bottleneck, and its benefit is therefore limited.
On XDNA~2, retaining $\boldsymbol{QK}^{\mathsf T}$ in MemTile memory is not sufficient to remove the MemTile bandwidth bottleneck. To move into the compute-bound regime, the dominant intermediate must be retained in compute-tile memory instead.
In that setting, a fused kernel becomes justified.

Early roofline analysis can identify when a mapping is already sufficiently fused for the target architecture, and thereby avoid unnecessary over-engineering.

\subsection{Use Broadcast to Exploit On-Chip Reuse}
\label{sec:lessons_broadcast}

Broadcast is an effective mechanism for exploiting on-chip reuse on spatial NPUs. When the same operand is required by multiple tiles, broadcast allows it to be transferred once and consumed by multiple destinations. This reduces communication volume and increases the OI.

In attention mappings, this pattern arises naturally because the same $\boldsymbol{Q}$, $\boldsymbol{K}$, or $\boldsymbol{V}$ blocks are often consumed by multiple tiles. All four strategies studied in this paper exploit broadcast for this purpose, and they all benefit from improved on-chip data reuse.

\subsection{Exploit Vector Parallelism}
\label{sec:lessons_hw}

The previous points focused on increasing OI, that is, moving the mapping rightward on the roofline. Once OI is sufficiently high, further performance depends on how efficiently the kernel uses the compute engine. On AMD NPUs, this requires exposing vector-level parallelism and utilizing the VLIW vector primitives whenever the computation permits.

On a VLIW NPU, available vector lanes should be used whenever the data layout and dependence structure allow it. Otherwise, the implementation may remain below the attainable roofline even when memory traffic is well optimized.

\section{Conclusion}
\label{sec:conclusion}

This paper presented a systematic ablation study of mapping FlashAttention onto AMD XDNA~1 and XDNA~2, evaluating four progressively optimized strategies: layer-by-layer execution, streamed dataflow, and fully fused kernel execution. Through roofline analysis at three memory hierarchy levels, we showed how the performance bottleneck shifts from shim-tile DMA bandwidth (layer-by-layer) through MemTile DMA bandwidth (DATO, IRON) to the compute-bound regime (MLIR-AIR).

The case study demonstrates that architecture-specific hardware-software co-design knowledge can be captured in reusable compiler flows while maintaining a high implementation-quality standard: the MLIR-AIR reference design reaches 3.62 TFLOP\textsubscript{GEMM}/s, 2.0$\times$ the evaluated IRON mapping, on XDNA~2.

On XDNA~2, the MLIR-AIR fused kernel achieves 2$\times$ speedup over the IRON pipeline, the state-of-the-art attention implementation on AMD NPUs. The implementation generalizes across LLM architectures---BERT, GPT-2, OPT, Llama (MHA and GQA), Qwen and DeepSeek---and with numerical stability evaluated for functional correctness.

From the ablation, we distilled actionable best practices for spatial NPU mapping.
These guidelines, while derived from attention on XDNA~2, apply broadly to memory-intensive multi-stage computations on spatial architectures. The roofline-driven ablation methodology provides a replicable and actionable framework for practitioners mapping similar workloads.

\bibliographystyle{ACM-Reference-Format}
\bibliography{references}

\end{document}